\documentstyle[12pt]{article}
\newcommand{\es}{\\[2mm]}

\newcommand{\journal}[4]{{\em #1~}#2\,(19#3)\,#4;}

\newcommand{\pr}{\journal {Phys. Rev.}}

\newcommand{\prl}{\journal {Phys. Rev. Lett.}}

\newcommand{\cmp}{\journal {Comm. Math. Phys.}}

\newcommand{\np}{\journal {Nucl. Phys.}}
\newcommand{\pl}{\journal {Phys. Lett.}}

\newcommand{\prep}{\journal {Phys. Reports}}

\newcommand{\nc}{\journal {Nuovo Cim.}}

\newcommand{\annp}{\journal {Ann. Phys. (N.Y.)}}

\makeatletter
\@addtoreset{equation}{section}
\makeatother

\def\Lp{\displaystyle{\biggl(}}
\def\Rp{\displaystyle{\biggr)}}
\def\LP{\displaystyle{\Biggl(}}
 \def\wti{\widetilde}
\def\RP{\displaystyle{\Biggr)}}
\newcommand{\lp}{\left(}\newcommand{\rp}{\right)}

\newcommand{\G}{\Gamma}
\newcommand{\D}{\Delta}

\renewcommand{\d}{\delta}
\newcommand{\e}{\varepsilon}

\newcommand{\g}{\gamma}

\renewcommand{\l}{\lambda} 
\newcommand{\m}{\mu}
\newcommand{\n}{\nu}

\newcommand{\p}{\psi}
\newcommand{\bp}{\bar \psi}
\newcommand{\r}{\rho}
\newcommand{\s}{\sigma} \renewcommand{\S}{\Sigma}

\newcommand{\WT}{\wti T}
\newcommand{\WS}{\wti \S}
\renewcommand{\AA}{{\cal A}}

\newcommand{\CC}{{\cal C}}

\newcommand{\PP}{{\cal P}}

\newcommand{\WW}{{\cal W}}

\newcommand{\complex}{{\kern .1em {\raise .47ex
\hbox {$\scriptscriptstyle |$}}
    \kern -.4em {\rm C}}}
\newcommand{\real}{{{\rm I} \kern -.19em {\rm R}}}
\newcommand{\rational}{{\kern .1em {\raise .47ex
\hbox{$\scripscriptstyle |$}}
    \kern -.35em {\rm Q}}}
\renewcommand{\natural}{{\vrule height 1.6ex width
.05em depth 0ex \kern -.35em {\rm N}}}
\newcommand{\dint}{\displaystyle{\int}}
\newcommand{\xint}{\dint d^4 \! x \, }

\newcommand{\pa}{\partial}

\newcommand{\fud}[2]  {{\displaystyle{\frac{\delta #1}{\delta #2}}}}
\newcommand{\dfrac}[2]{{\displaystyle{\frac{#1}{#2}}}}

\newcommand{\sla}{\raise.15ex\hbox{$/$}\kern -.57em}

\newcommand{\twiddle}{\lower.9ex\rlap{$\kern -.1em\scriptstyle\sim$}}

\newcommand{\equ}[1]{(\ref{#1})}

\newcommand{\eq}{\begin{equation}}
\newcommand{\eqn}[1]{\label{#1}\end{equation}}
\newcommand{\eea}{\end{eqnarray}}
\newcommand{\eqa}{\begin{eqnarray}}
\newcommand{\eqan}[1]{\label{#1}\end{eqnarray}}
\newcommand{\ba}{\begin{array}}
\newcommand{\ea}{\end{array}}
\newcommand{\eqac}{\begin{equation}\begin{array}{rcl}}
\newcommand{\eqacn}[1]{\end{array}\label{#1}\end{equation}}

\begin{document}
{\large     


{\ }

\vspace{20mm}

\vspace{2cm}
\centerline{\LARGE Algebraic renormalization of antisymmetric}  \vspace{2mm}

\centerline{\LARGE tensor matter fields }  \vspace{2mm}

\vspace{9mm}

\centerline{ Vitor Lemes, Ricardo Renan}
\centerline{ and }
\centerline{S.P. Sorella}
\vspace{4mm}
\centerline{\it C.B.P.F}
\centerline{\it Centro Brasileiro de Pesquisas Fisicas,}
\centerline{\it Rua Xavier Sigaud 150, 22290-180 Urca}
\centerline{\it Rio de Janeiro, Brazil}
\vspace{10mm}

\centerline{{\normalsize {\bf PACS: 11.15.Bt }} }
\vspace{4mm}
\centerline{{\normalsize {\bf REF. 032/94}} }

\vspace{4mm}
\vspace{10mm}

\centerline{\Large{\bf Abstract}}\vspace{2mm}
\noindent
The algebraic renormalization of a recently proposed abelian axial
gauge model with antisymmetric tensor matter fields is presented.

\setcounter{page}{0}
\thispagestyle{empty}

\vfill
\pagebreak

\section{Introduction}

Antisymmetric tensor fields have been introduced since many
years~\cite{ant} and
are object of continuous and renewed interests due to their connection
with the topological field theories~\cite{bbrt}. They are indeed the building
blocks of a large class of Schwarz type topological theories~\cite{schw}.
These
theories, also known as $BF$ models, can be formulated in any
space-time dimension and allow to compute topological invariants which
generalize the three dimensional linking number~\cite{link}.
Moreover they provide
an example of ultraviolet finite field theories~\cite{uf}.
Let us point out that
the antisymmetric tensor fields of the topological models are gauge
fields, i.e. they possess zero modes which have to be gauge fixed in
order to have a nondegenerate action.

More recently L. V. Avdeev and M. V. Chizhov~\cite{ac}
achieved the construction
of a four dimensional abelian gauge model which includes antisymmetric
second rank tensor fields as matter fields rather than gauge fields.
The model contains also a coupling of the antisymmetric fields with
chiral spinors and a quartic tensor self-interaction term. It exhibits
several interesting features among which we underline the asymptotically
free ultraviolet behaviour of the abelian gauge interaction. As shown
by the authors~\cite{ac} with an explicit one-loop computation, this is
due to
the fact that the contribution of the tensor fields to the gauge
$\beta$-function is negative. Antisymmetric matter fields represent thus
a possibility of introducing a new type of interaction in gauge theories.

The aim of this work is to give a regularization independent
algebraic analysis of the model. The use
of the algebraic method~\cite{alg} in the present case is motivated by
the fact
that, besides the presence of $\g_5$ in the axial couplings of the
classical action, the gauge transformations as well as the interaction
vertices involving the antisymmetric matter fields explicitely contain
the $\e_{\m\n\rho\sigma}$ tensor.

Moreover
one has to be sure that the introduction of a new type of matter
field does not lead to a new kind of anomaly. Indeed, as we shall see
by exploiting the Wess-Zumino consistency condition~\cite{wz}, the tensor
field gives rise to a nontrivial cocycle which is
independent from the gauge field and which contains only matter fields.
Its numerical coefficient turns out to be related to a set of one-loop
Feynman diagrams which are built-up with the interaction vertex between
the tensor and the spinor fields. However we will be able to
show that the tensor-spinor vertex as well as the matter cocycle can be
consistently eliminated by requiring that the model is invariant under
an additional descrete symmetry. One is left thus with the usual gauge
anomaly of the chiral $QED$. Furthermore the Adler-Bardeen theorem
ensures that the latter is definitively absent if it is absent at the
one-loop level, guaranteing then the ultraviolet renormalizability of
the model.

The paper is organized as follows. In Sect. 2 we introduce the classical
action and we discuss its stability under radiative corrections.
Sect. 3 is devoted to the analysis of the Wess-Zumino consistency
condition. In Sect. 4 we present a geometrical derivation of the
model in terms of covariant derivatives. Finally, Sect. 5 contains
some useful relations and the conventions.

\section{The model and its stability}

The model is specified by the following classical action which, using the
same notation of ref.~\cite{ac}, reads
\eq\ba{rl}
S_{inv} = \xint \LP  & -\dfrac{1}{4h^2} F_{\m\n} F^{\m\n} +
        i \bp \g^{\m} \pa_{\m} \p - \bp \g^{\m}\g_{5}A_{\m}\p
   +  \dfrac{1}{2} {(\pa_{\l}T_{\m\n})}^2  \es
&  - 2 {(\pa^{\m}T_{\m\n})}^2
  + 4 A_{\m}\lp T^{\m\n}\pa^{\l}\WT_{\l\n} -
             \WT^{\m\n}\pa^{\l} T_{\l\n} \rp  \es
&   + 4 \lp \dfrac{1}{2} {(A_{\l}T_{\m\n})}^2 -2 {(A^{\m}T_{\m\n})}^2 \rp
    + y \bp \s_{\m\n} T^{\m\n} \p   \es
&   + \dfrac{q}{4} \lp \dfrac{1}{2}{(T_{\m\n}T^{\m\n})}^2 -
               2 T_{\m\n}T^{\n\r}T_{\r\l}T^{\l\m} \rp {\ } \RP \ ,
\ea\eqn{inv-actions}
where $(h,y,q)$ are coupling constants and $T_{\m\n}=-T_{\m\n}$ is a
second rank antisymmetric tensor field with
\eq\ba{l}
     \WT_{\m\n} = \dfrac{1}{2}\e_{\m\n\r\s}T^{\r\s} \ ,  \es
     \dfrac{1}{2}\e_{\m\n\r\s}\WT^{\r\s} = - T_{\m\n} \ ,
\ea\eqn{tildeT}
the $\e_{\m\n\r\s}$ tensor being normalized as
\eq
    \e_{\m_{1}\m_{2}\m_{3}\m_{4}} \e^{\n_{1}\n_{2}\n_{3}\n_{4}} =
      - \d^{[\n_{1}}_{\m_{1}}... \d^{\n_{4}]}_{\m_{4}}  \ .
\eqn{eps-normal}
We use the Minkowski metric $\eta_{\m\n}= {\rm diag}(+---)$ and we
denote $\s^{\m\n}=\frac{i}{2}[\g^{\m},\g^{\n}]$ (see Sect. 5 for the
conventions).

Notice that the quadratic term in the antisymmetric field of expression
\equ{inv-actions} is nondegenerate, i.e. $T_{\m\n}$ is a matter field. This
feature allows also for the introduction of a spinor-tensor interaction
($y$-term) as well as for a quartic tensor self-interaction ($q$-term).
It is easy to check that the action \equ{inv-actions} is left invariant
by the following gauge transformations~\cite{ac}:
\eq\ba{l}
\d A_\m = \pa_\m\omega  \ , \es
\d \p = -i\omega \g_{5} \p \ , \es
\d \bp = -i\omega \bp \g_{5} \ , \es
\d T_{\m\n} = -2\omega \WT_{\m\n} \ .
\ea\eqn{gauge-transf}
Introducing now a covariant Feynman gauge
\eq
  S_{gf} =  -\dfrac{1}{2\alpha} \xint   {(\pa A)}^2  \ ,
\eqn{gauge-f}
the gauge-fixed action
\eq
\S = S_{inv} + S_{gf} \ ,
\eqn{g-f-action}
obeys the Ward identity
\eq
 \WW(x) \S = -\dfrac{1}{\alpha} \pa^2 \pa A \ ,
\eqn{ward-id}
where $\WW(x)$ denotes the local Ward operator\footnote{As usual the spinor
derivative $\d \over {\d \p}$ acts from the right to the left.}
\eq
\WW(x) = -\pa_{\m}\fud{{\ }}{A_{\m}} - i\fud{{\ }}{\p} \g_{5}\p
         -i\bp \g_{5}\fud{{\ }}{\bp} - \WT_{\m\n}\fud{{\ }}{T_{\m\n}} \ .
\eqn{ward-op}
The classical action $\S$ is known to be constrained, besides the Ward
identity \equ{ward-id}, by a set of descrete symmetries: i.e. parity $\PP$
and charge conjugation $\CC$~\cite{oliv,itz}. They act as

\vspace{4mm}
${\ }{\ }i)$ Parity $\PP$
\eq\ba{l}
 x \rightarrow  x_p=(x^0,-x^{i})  \ , \qquad i=1,2,3  \es
 \p(x_{p}) = \g^0 \p(x) \ , \es
 A_0(x_{p}) = -  A_0(x) \ ,  \qquad  A_i(x_{p}) =  A_i(x) \ , \es
 T_{0i}(x_{p}) = -  T_{0i}(x) \ , \qquad  T_{ij}(x_{p}) =  T_{ij}(x)  \ .
\ea\eqn{parity}

\vspace{4mm}
${\ }{\ }ii)$ Charge conjugation $\CC$
\eq\ba{l}
  \p \rightarrow \p^{c} = C \bp^{T} \ , \qquad C=i\g^{0} \g^{2} \ , \es
  A_{\m} \rightarrow  A_{\m}^{c} =  A_{\m}  \ , \es
  T_{\m\n} \rightarrow  T_{\m\n}^{c} = - T_{\m\n}  \ .
\ea\eqn{c-conjug}
The fields $(A,\p,T)$ have respectively dimension $(1,3/2,1)$.

In order to study the stability~\cite{alg} of the model under radiative
corrections
we look at the most general solution of the equation
\eq
  \WW(x) \WS = 0 \ ,
\eqn{stab}
where $\WS$ is an integrated local polynomial in the fields and their
derivatives with dimension four, invariant under parity $\PP$ and charge
conjugation $\CC$. $\WS$ represents the most general local gauge invariant
counterterm which one can freely add at each order of perturbation theory.
It can be parametrized as
\eq
 \WS = \WS(A,\p) + \WS(A,\p,T) \ ,
\eqn{par-countert}
where $\WS(A,\p)$ depends only on the fields $(A,\p,\bp)$ and
$\WS(A,\p,T)$ collects the dependence on the new tensor matter field
$T_{\m\n}$. The stability condition \equ{stab}, due to the fact that the
Ward operator $\WW(x)$ is linear, splits into the two conditions
\eq
  \WW(x) \WS(A,\p) = 0 \ ,
\eqn{stab1}
and
\eq
  \WW(x) \WS(A,\p,T) = 0 \ .
\eqn{stab2}
The first equation \equ{stab1} yields the well known local invariant
counterterm of the chiral $QED$~\cite{oliv}
\eq
\WS(A,\p) =  \xint \LP   -\dfrac{\r}{4} F_{\m\n} F^{\m\n} +
      \s \lp  i \bp \g^{\m} \pa_{\m} \p - \bp \g^{\m}\g_{5}A_{\m}\p \rp \RP \ ,
\eqn{count1}
with $(\r,\s)$ arbitrary parameters. Turning now to the second term
$\WS(A,\p,T)$ it follows that, using the algebraic identity valid in four
dimensions
\eq
\e^{\alpha\beta\m\n}\pa^{\s} + \e^{\s\alpha\beta\m}\pa^{\n} +
\e^{\n\s\alpha\beta}\pa^{\m} + \e^{\m\n\s\alpha}\pa^{\beta} +
\e^{\beta\m\n\s}\pa^{\alpha} = 0  \ ,
\eqn{four-ident}
it can be parametrized as
\eq\ba{rl}
\WS(A,\p,T) = \xint \LP  &  a {(T_{\m\n}T^{\m\n})}^2 +
    b  T_{\m\n}T^{\n\r}T_{\r\l}T^{\l\m} + c  {(\pa_{\l}T_{\m\n})}^2  \es
 &  + d {(\pa^{\m}T_{\m\n})}^2  + e \pa A T_{\m\n}\WT^{\m\n}
    + m  A_{\m} T^{\m\n}\pa^{\l}\WT_{\l\n}   \es
 &  + n A_{\m}\WT^{\m\n}\pa^{\l} T_{\l\n}  + p {(A^{\m}T_{\m\n})}^2  \es
 &  + u {(A_{\m}T_{\l\n})}^2   +  v \bp \s_{\m\n} T^{\m\n} \p  \RP \ ,
\ea\eqn{count2}
with $(a,b,c,d,e,m,n,p,u,v)$ constant parameters.
Condition \equ{stab2} implies that $\WS(A,\p,T)$ depends only on three
parameters, i.e.
\eq\ba{rl}
\WS(A,\p,T) = & \xint \LP{\ }  c \LP
   \dfrac{1}{2} {(\pa_{\l}T_{\m\n})}^2  -2 {(\pa^{\m}T_{\m\n})}^2
  + 4 A_{\m} T^{\m\n}\pa^{\l}\WT_{\l\n} \es
  &{\ }{\ } -4 A_{\m} \WT^{\m\n}\pa^{\l} T_{\l\n}
 + 4 \Lp\dfrac{1}{2} {(A_{\m}T_{\l\n})}^2 - 2 {(A^{\m}T_{\m\n})}^2 \Rp \RP \es
  & {\ }{\ }+ a \Lp \dfrac{1}{2} {(T_{\m\n}T^{\m\n})}^2 -
       2 T_{\m\n}T^{\n\r}T_{\r\l}T^{\l\m} \Rp
      +  v \bp \s_{\m\n} T^{\m\n} \p  \RP \ .
\ea\eqn{count2r}
One sees thus that the most general local gauge invariant counterterm
contains five free independent parameters $(\r,\s,c,a,v)$. They are easily
seen to correspond to a renormalization of the coupling constants
$(h,y,q)$ and to field amplitude redefinitions\footnote{As it happens in
the ordinary $QED$, the gauge fixing term \equ{gauge-f} is not renormalized.}.
This proves the stability of the classical action under radiative
corrections.

\section{The Wess-Zumino consistency condition}

At the quantum level the action $\S$ is replaced by a vertex functional
\eq
  \G = \S + O(\hbar) \ ,
\eqn{vertex-funct}
which obeys the broken Ward identity
\eq
 \WW(x) \G = -\dfrac{1}{\alpha} \pa^2 \pa A  + \AA(x)\cdot\G  \ ,
\eqn{broken-ward-id}
where the insertion $\AA\cdot\G$ represents the possible breaking induced
by the radiative corrections. According to the
Quantum Action Principle~\cite{qap}
the lowest order nonvanishing contribution to the breaking - of order
$\hbar$ at least -
\eq
  \AA\cdot\G = \AA + O(\hbar\AA) \ ,
\eqn{first-order}
is a local functional with dimension four, even under charge conjugation
$\CC$, odd under parity $\PP$\footnote{This property follows from the
fact that the Ward operator $\WW(x)$ \equ{ward-op} is  odd under parity
transformations and even under charge conjugation.} and constrained
by the Wess-Zumino consistency condition~\cite{wz}
\eq
 \WW(x) \AA(y){\ }-{\ } \WW(y)\AA(x) = 0 \ .
\eqn{wess-zumino}
The latter stems from the algebraic relation
\eq
  \WW(x) \WW(y){\ }-{\ }\WW(y) \WW(x) = 0 \ .
\eqn{alg-relww}
As it is well known the theory will be anomaly free if condition
\equ{wess-zumino} admits only the trivial solution, i.e.
\eq
 \AA^{tr}(x)= \WW(x)\D \ ,
\eqn{triv-sol}
with $\D$ an integrated local polynomial with dimension four and even
under parity $\PP$ and charge conjugation $\CC$. On the
other hand nontrivial cocycles of \equ{wess-zumino}
\eq
  \AA^{nontr}(x) \ne \WW(x)\D \ ,
\eqn{ntriv-sol}
cannot be reabsorbed as local counterterms and represent real obstructions
in order to have an invariant quantum vertex functional.

To study the Wess-Zumino consistency condition we proceed as before
and we write
\eq
 \AA(x) = \AA_{1}(A,\p) + \AA_{2}(A,\p,T)  \ ,
\eqn{decanom}
where $\AA_{1}(A,\p)$ depends only on $(A,\p,\bp)$ and $\AA_{2}$ contains
the field $T_{\m\n}$. Equation \equ{wess-zumino} splits thus into the
two conditions
\eq
 \WW(x) \AA_{1}(y){\ }-{\ } \WW(y)\AA_{1}(x) = 0 \
\eqn{wess-zumino1}
and
\eq
 \WW(x) \AA_{2}(y){\ }-{\ } \WW(y)\AA_{2}(x) = 0 \ .
\eqn{wess-zumino2}
The first equation \equ{wess-zumino1} yields, modulo trivial cocycles,
the usual abelian gauge anomaly~\cite{anom}
\eq
 \AA_{1} = r \e_{\m\n\r\s} \pa^{\m} A^{\n} \pa^{\r} A^{\s} \ .
\eqn{abelian-anom}
Turning to the second equation \equ{wess-zumino2} it is not difficult to
show the existence of a nontrivial $T_{\m\n}$-dependent cocycle. It
reads
\eq
 \AA_{2} = \eta \bp \s_{\m\n}\g_{5} \p T^{\m\n} \ ,
\eqn{t-anom}
with $\eta$ a numerical coefficient.
Notice that expression \equ{t-anom} depends only on the matter fields
and does not contain the gauge field $A$. Acting now on the Ward identity
\equ{broken-ward-id} with the test operator
\eq
   \fud{{\ }}{\p(y)} \fud{{\ }}{\bp(z)} \fud{{\ }}{T_{\m\n}(u)} \ ,
\eqn{testop}
and setting all the fields equal to zero, one easily checks that the
numerical coefficient $\eta$, at its lowest order, is related to two kinds
of one-loop $1PI$ diagrams $\G_{A\p\bp T}^{(1)}$ and $\G_{\p\bp T}^{(1)}$,
respectively a set of three box diagrams with four amputated external
legs of the type $(A,\p,\bp,T)$ and three triangle diagrams with
$(\p,\bp, T)$ as external amputated legs. The appearence of the matter cocycle
$\bp \s_{\m\n}\g_{5} \p T^{\m\n}$ could jeopardize the Adler-Bardeen
theorem~\cite{a-b,proof} of the gauge anomaly and spoil the
renormalizability of the theory.

One is forced then to impose further constraints on the model in order
to avoid the presence of the cocycle \equ{t-anom}. To this purpose let
us remark that all the diagrams which contribute to the coefficient
$\eta$ are built-up by making use of the tensor-spinor three vertex
$(\p\bp T)$ of \equ{inv-actions}. Therefore the simplest mechanism which
one may impose in order to exclude the matter cocycle is to require
from the very beginning that the classical action \equ{inv-actions} is
invariant under an additional descrete symmetry
\eq
    T_{\m\n} \rightarrow - T_{\m\n}  \ .
\eqn{descrte-sym}
It is apparent that this new invariance forbids the presence in
\equ{inv-actions} of the three vertex $(\p\bp T)$. The latter, as it
follows by combining \equ{descrte-sym} with the stability analysis of
Sect. 2, will be not reintroduced by the radiative corrections.
In addition, requirement \equ{descrte-sym} excludes the appearence
of \equ{t-anom} as an anomaly in the Ward identity \equ{broken-ward-id}.
We are left then with the gauge anomaly \equ{abelian-anom}.
However, the Adler-Bardeen theorem~\cite{proof} ensures that if the coefficient
$r$ vanishes at one-loop order it will vanish at all orders of
perturbation theory, guaranteing thus the perturbative renormalizability
of the model. As it is well known~\cite{oliv}
(see also the detailed discussion of the authors~\cite{ac}), the vanishing of
the coefficient $r$ at the one-loop
order in the present abelian case is achieved by introducing a partner for
every charged particle with opposite axial gauge charge.

Of course, the descrete symmetry \equ{descrte-sym} will extend to the
partner of the tensor matter field $T_{\m\n}$.

\section{A geometrical construction}

In this section we present a simple geometrical derivation of the model.
Let us begin by introducing the {\it covariant derivative}
\eq
  \nabla_{\m} T_{\r\s} := (\pa_{\m}\WT_{\r\s} - 2 A_{\m}T_{\r\s} ) \ ,
\eqn{cov-deriv}
and
\eq
  {\nabla_{\m} \WT_{\r\s}} = \dfrac{1}{2}
          \e_{\r\s\alpha\beta}\nabla_{\m}T^{\alpha\beta} =
         -(\pa_{\m} T_{\r\s} + 2 A_{\m}\WT_{\r\s} ) \ .
\eqn{cov-deriv-eps}
The name {\it covariant} is justified by the fact that $(\nabla T)$
and $(\nabla \WT)$ transform under the gauge transformations
\equ{gauge-transf} as the fields $T$ and $\WT$, i.e.
\eq\ba{l}
\d (\nabla_{\m}T_{\r\s}) = -2\omega (\nabla_{\m}\WT_{\r\s}) \ , \es
\d (\nabla_{\m}\WT_{\r\s}) = 2\omega (\nabla_{\m}T_{\r\s}) \ .
\ea\eqn{nabla-gauge-transf}
This important property allows to construct gauge invariant quantities
in a very simple way. It is almost immediate indeed to verify that the term
\eq
 {\tilde S} = \xint \LP (\nabla^{\m}T_{\m\n}) (\nabla^{\s}T_{\s}^{{\ }\n})
        +  (\nabla^{\m}\WT_{\m\n}) (\nabla^{\s}\WT_{\s}^{{\ }\n}) \RP \ ,
\eqn{gauge-inv}
is gauge invariant.

Observe that a different contraction of the Lorentz indices
\eq
 \xint \LP (\nabla_{\m}T_{\r\n}) (\nabla^{\m}T^{\r\n})
        +  (\nabla_{\m}\WT_{\r\n}) (\nabla^{\m}\WT^{\r\n}) \RP \ ,
\eqn{gauge-inv-zero}
identically vanishes due to the property
\eq
 \WT_{\r\s} \WT^{\r\s} = - T_{\r\s} T^{\r\s}  \ .
\eqn{tt-relation}
Expression \equ{gauge-inv} is easily computed to be
\eq\ba{rl}
 {\tilde S} = - \xint \LP &
   \dfrac{1}{2} {(\pa_{\l}T_{\m\n})}^2  -2 {(\pa^{\m}T_{\m\n})}^2
  + 4 A_{\m} T^{\m\n}\pa^{\l}\WT_{\l\n} \es
  & -4 A_{\m} \WT^{\m\n}\pa^{\l} T_{\l\n}
 + 4 \Lp \dfrac{1}{2} {(A_{\m}T_{\l\n})}^2 - 2 {(A^{\m}T_{\m\n})}^2 \Rp \RP \ ,
\ea\eqn{stilde-expr}
i.e. $\tilde S$ is nothingh but the $(A-T)$ dependent part of the expression
\equ{inv-actions}. We have thus recovered in an elegant and more geometrical
way the initial invariant classical action $S_{inv}$. The generalization of the
above construction for a nonabelian version of \equ{inv-actions} is under
investigation.

\section{Conventions}

We give here some useful relations involving the $\e_{\m\n\r\s}$ and the
Dirac $\g$-matrices.

We have
\eq\ba{l}
\WT_{\m\l} \WT^{\l\n} = T_{\m\l} T^{\l\n} + \dfrac{1}{2} \d^{\n}_{\m}
                         T_{\alpha\beta} T^{\alpha\beta} \ , \es
T_{\m\l}\WT^{\l\n} = \dfrac{1}{4} \d^{\n}_{\m}
                    T_{\alpha\beta} \WT_{\alpha\beta} \ .
\ea\eqn{conv1}
{}From equation \equ{four-ident} one gets
\eq\ba{l}
(\pa_{\s}\WT_{\m\r}) \WT^{\s\r} = -\dfrac{1}{2}
          T^{\alpha\beta} \pa_{\m} T_{\alpha\beta}
     - T_{\m\r} \pa^{\s}T^{\r}_{\s}  \ , \es
 \WT_{\m\r}\pa_{\s} \WT^{\s\r} = -\dfrac{1}{2}
          T^{\alpha\beta} \pa_{\m} T_{\alpha\beta}
     - (\pa_{\s}T_{\l\m}) T^{\s\l} \ ,
\ea\eqn{conv2}
as well as
\eq
\WT_{\alpha\beta} \pa^2 T^{\alpha\beta} =
  -2 T_{\alpha\beta}\pa_{\n}\pa^{\beta} \WT^{\n\alpha}
  -2 \WT_{\alpha\beta} \pa_{\n} \pa^{\beta} T^{\n\alpha} \ .
\eqn{conv3}
Concerning the $\g$-matrices, here taken in the Dirac representation, we
use~\cite{ac,itz}:
\eq
 \g_{5} = i \g^{0} \g^{1} \g^{2} \g^{3} \ , \qquad  {\g_{5}}^2 =1 \ ,
\eqn{conv4}
\eq
 \s^{\m\n} = \dfrac{i}{2}[\g^{\m},\g^{\n}] \ , \qquad
    \dfrac{i}{2} \e^{\m\n\alpha\beta}\s_{\alpha\beta} = \g_{5} \s^{\m\n} \ .
\eqn{conv5}
For the charge conjugation matrix $C$
\eq
  C = i \g^{0} \g^{2} \ , \qquad C^2=-1 \ , \qquad C^{-1} = -C \ ,
\eqn{charge-conj-matrix}
and
\eq
  C\g_{\m} C = \g_{\m}^{T} \ , \qquad
  C\g_{5} C = - \g_{5} = - \g_{5}^{T}  \ .
\eqn{conv6}

\vspace{2cm}

\noindent{\large{\bf Acknowledgements}}

We wish to thank Jos{\'e} Helayel-Neto for a carefully reading of the
manuscript.
The {\it Conselho Nacional de Desenvolvimento Cientifico e Tecnologico},
$CNPq$-Brazil is gratefully acknowledged for the financial support.


}    
\end{document}